\documentclass[preprint]{ptephy_v1}

\preprintnumber{XXXX-XXXX} 

\usepackage{hyperref} 
\usepackage{url} 
\usepackage{lineno}
\usepackage{setspace}
\begin{document}
\title{Improvement of radon detector performance by using a large-sized PIN-photodiode}


\author[1]{K.~Okamoto}
\author[1,*]{Y.~Nakano}
\author[1]{G.~Pronost}
\author[1,2]{H.~Sekiya}
\author[1]{S.~Tasaka}
\author[3,2]{Y.~Takeuchi}
\author[1,2]{M.~Nakahata\thanks{These authors contributed equally to this work}}

\affil[1]{Kamioka Observatory, Institute for Cosmic Ray Research, the University of Tokyo, Gifu 506-1205, Japan}
\affil[2]{Kavli Institute for the Physics and Mathematics of the Universe~(WPI), the University of Tokyo Institutes for Advanced Study, the University of Tokyo, Kashiwa, Chiba 277-8583, Japan}
\affil[3]{Department of Physics, Graduate School of Science, Kobe University, Kobe, Hyogo 657-8501, Japan \email{ynakano@km.icrr.u-tokyo.ac.jp}}


\begin{abstract}%
Radioactive noble gas radon~($\mathrm{^{222}Rn}$) is one of the major background sources below the MeV region in rare event search experiments. To precisely measure radon concentration in purified gases, a radon detector with an electrostatic collection method is widely used. In this paper, we discussed the improvements of a radon detector by installing a new PIN-photodiode~($28\times28~\mathrm{mm}$) whose surface area is $2.5$~times larger than that used previously~($18\times18~\mathrm{mm}$). We evaluated the detector's performance by serially connecting two radon detectors equipped with two types of PIN-photodiodes. As a result of the calibrations, we found an improvement of $(3.8\pm2.4)\%$ in the detection efficiencies below $1.0~\mathrm{g/m^{3}}$, while a $10$--$20\%$ improvement occurred above this level. The intrinsic background of the detector equipped with the large PIN-photodiode was measured as $0.24^{+0.09}_{-0.05}~\mathrm{mBq/m^{3}}$. This background level is consistent with the radon detector with the small PIN-photodiode, although we installed the large one. This improvement is useful for applications in radon emanation measurements from a material, which also emits water from its surface.
\end{abstract}

\subjectindex{xxxx, xxx}

\maketitle

\section{Introduction} \label{sec_intro}

Noble gas radon~($\mathrm{^{222}Rn}$, hereinafter Rn) is one of the major background sources in experiments that search for rare~(interaction/decay) events, such as solar neutrinos, neutrino-less double beta decay, and dark matter. Once a detector for a particular experiment is constructed, Rn gas is continuously generated from the decay chain of the $\mathrm{^{238}U}$ series in the detector material. Owing to its long half-life~($3.82$~days), its diffusion into the detector's fiducial volume results in a mimicking signal in that energy range. Therefore, material screening before the detector construction, monitoring Rn in target materials used in such experiments, and Rn removal during the detector operation by their purification processes are essential for improving their sensitivities.

For many years, several technologies for Rn measurements have been developed for these purposes. In particular, a Rn detector equipped with a PIN-photodiode is widely used to measure the Rn concentration in gases~\cite{4329405, HOWARD1990589, 1994125, 1996741, 1997710, TAKEUCHI1999334, CHOI2001177, KIKO2001272, Mitsuda:2003eu, GUTIERREZ2004583, MARTINMARTIN20061287, Mamedov:2011zz, 2013InJPh..87..471A, Hosokawa:2015koa, Nakano:2017rsy, Pronost:2018ghn, Elisio:2019nuz}. The principal technique of such Rn detectors is electrostatic collection~\cite{Kotrappa1981ElectretaNT}. Because the daughter nuclei of $\mathrm{^{222}Rn}$ tend to be positively charged~\cite{PMID:2663781}, they are easily collected by forming an electrical field inside the detector. The PIN-photodiode, where charged nuclei are collected, uniquely identifies the energy of the $\alpha$ particle emitted via the daughter's subsequent decays because of its high spectral resolution.

For the past $30$~years, we have developed this type of Rn detector. The current design of this detector, whose inner volume is approximately $80$~litter~(hereinafter~L), has been used to measure Rn concentrations less than a few $\mathrm{mBq/m^{3}}$ in several gases~\cite{Hosokawa:2015koa, Nakano:2017rsy}. Recently, we applied this detector to measure the Rn concentration in purified water in the Super-Kamiokande detector by combining it with a Rn extraction column and cooled charcoal~\cite{Nakano:2019bnr} and to evaluate several absorbents for removing Rn from purified gases~\cite{Ogawa:2019ccp, Nakano:2020tej}.

For future large-scale experiments to search for rate events, further sensitivity is required to monitor the Rn concentration below the $\mu\mathrm{Bq/m^{3}}$ level in purified gases~\cite{Abe:2011ts, Aalseth:2017fik, Aprile:2020vtw, Agostini:2020adk}. There are four strategies for improving the detection sensitivity of such Rn detectors. The first is to enlarge the detector volume, the second is to reduce the intrinsic background, the third is to raise the supplying high voltage to form a large electric field inside the detector, and the fourth is to enlarge the surface area of the PIN-photodiode. The details of these strategies are as follows.

\begin{description}
\item[1.~Detector volume] \mbox{}\\
The enlargement of the detector volume is an easy way to improve the sensitivity. For instance, to measure Rn concentrations less than a few $\mathrm{mBq/m^{3}}$ in gases, several large Rn detectors have been developed in the Borexino collaboration~($480$~L~\cite{Simgen:2003et}), Super-Kamiokande collaboration~($700$~L~\cite{Mitsuda:2003eu}), and JUNO collaboration ($279$~L~\cite{Yu:2020vbz}). On the other hand, this strategy has two disadvantages: it is cost-intensive and increases the intrinsic background emanating from the inner surface of the detector structure.
\item[2.~Intrinsic background] \mbox{}\\
The background of such a detector comes from Rn emanating from the inner surface as well as air leaks. Electropolishing the inner surface efficiently reduces Rn emanating from detector materials~\cite{TAKEUCHI1999334}. Furthermore, the latest $80$~L detector used knife-edge flanges with metal gaskets~(CF flanges) to prevent the entry of external air~\cite{Hosokawa:2015koa, Nakano:2017rsy}. This technique enables the detector to be airtight. Based on these techniques, the background rate of the latest $80$~L Rn detector was evaluated as $0.74\pm0.07~\mathrm{(Statistical~uncertainty~only)}$~count/day when purified air was filled with~\cite{Nakano:2017rsy}. Owing to its low background rate, the sensitivity was less than $0.5~\mathrm{mBq/m^{3}}$ level by a single-day measurement.

Another possibility is copper electroplating because the radio-contaminants of copper materials are generally low~\cite{hoppe2008, Alduino:2016vjd, Abe:2017jzw, Bunker:2020sxw}. This technique may reduce the intrinsic background from the inner surface of the detector because the metal layer on its surface acts as an internal shield~\cite{Hoppe:2014nva, Balogh:2020nmo}. However, this approach, as well as the enlargement of the detector volume, is expensive.
 
\item[3.~High voltage] \mbox{}\\
High voltage directly affects the collection efficiency in principle because the charged nuclei are collected by the electrical field formed between the PIN-photodiode and a stainless steel structure. Therefore, increasing the supplied voltage is an easy way to improve the sensitivity of the detector. However, this approach depends on the shape of the detector and the location of the installed PIN-photodiode. In other words, the sensitivity of the detector is limited by the detector's size because of the difficulty in forming the electrical field efficiently inside the detector.

\item[4.~PIN-photodiode] \mbox{}\\
The surface area of the PIN-photodiode is sensitive to the collection efficiency because the charged nuclei are eventually collected in that area. Furthermore, the electrical field formed by the small size of the PIN-photodiode is screened by the surrounding stainless structure~\cite{2011JInst...6C2018F} as the electrical field formation is determined by an electrical dipole. Therefore, a larger surface area enables the Rn detector to more efficiently collect the charged nuclei and increase the volume where the electrical field is applied inside the detector.
\end{description}

Based on these discussions, we chose a PIN-photodiode whose surface area is $2.5$~times larger than that used previously as a first step. This approach enables the detector to improve its sensitivity at a low cost, while other approaches are cost-intensive.

This article consists of six sections, including an introduction. In Sect.~\ref{sec_detector}, we briefly introduce the developed Rn detector and describe its characteristics. In addition, we discuss the advantages of installing a large PIN-photodiode to the Rn detector. In Sect.~\ref{sec_setup}, we describe the setup of and results from the calibration using purified air. In particular, we report the high voltage and absolute humidity dependence of the collection efficiency and compare the results with those obtained in previous studies. In Sect.~\ref{sec_bg}, we evaluate the background of the Rn detector and compare it with past results. In Sect.~\ref{sec_future}, we discuss the applications of the Rn detector in precise measurement programs. Finally, we summarize and conclude the study in Sect.~\ref{sec_conclusion}.

\section{Rn detector} \label{sec_detector}
\subsection{Detector structure and electronics}
\begin{figure*}
\centering\includegraphics[width=0.9\linewidth]{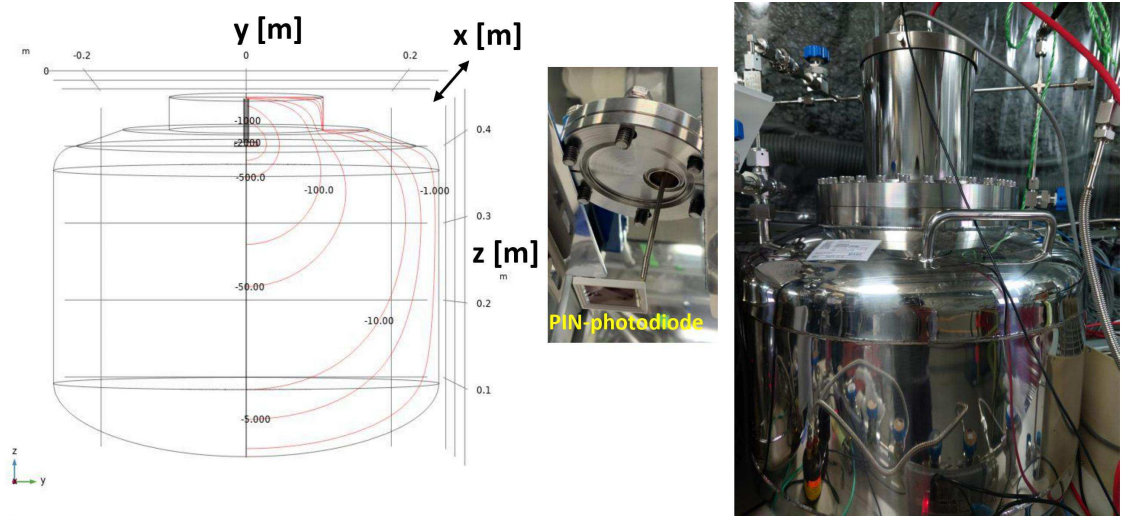}
\caption{Left: The conceptual diagram of the Rn detector and the electrical field simulated by COMSOL Multiphysics$^{\textcircled{\footnotesize{R}}}$. In this simulation, $28\times28~\mathrm{mm}$ of the PIN-photodiode is installed at the top of the detector, and $-2.0$~kV is supplied to form the electrical field. The red solid lines show the equipotential lines~($-2000$~V, $-1000$~V, $-500$~V, $-100$~V, $-50$~V, $-10$~V, $-5$~V, and $-1.0$~V, respectively). Right: Image of the Rn detector and the PIN-photodiode.}
\label{fig_det}
\end{figure*}

The Rn detector consists of a stainless steel vessel, a PIN-photodiode, a ceramic feed through, and an electrical circuit. Figure~\ref{fig_det} shows the structure of the $80$~L Rn detector and the electrical field simulated by COSMOL Multiphysics$^{\textcircled{\footnotesize{R}}}$\footnote{\url{https://www.comsol.jp/comsol-multiphysics}}. The PIN-photodiode was connected to the inner end of the feedthrough, which was mounted on the top flange of the vessel. The high-voltage divider and readout electronics are also connected to the external end of the feedthrough. The details of the Rn detector can be found in Ref.~\cite{Hosokawa:2015koa, Nakano:2017rsy}.

The electronics consisted of a high-voltage supplier and an amplifier circuit. The reverse bias voltage was simultaneously supplied to the PIN-photodiode. In this study, we used a Raspberry Pi-based system originally developed for monitoring the Rn concentration in the underground experimental area in the Kamioka mine with a $1$~L Rn detector~\cite{Pronost:2018ghn}. The data acquisition system recorded the signal from the Rn detector and the environmental parameters, such as the temperature, dew point, inner pressure of the detector, and flow rate of gas circulation in the calibration system.

\subsection{Larger size PIN-photodiode and formed electrical field}
\begin{figure}
\centering\includegraphics[width=0.8\linewidth]{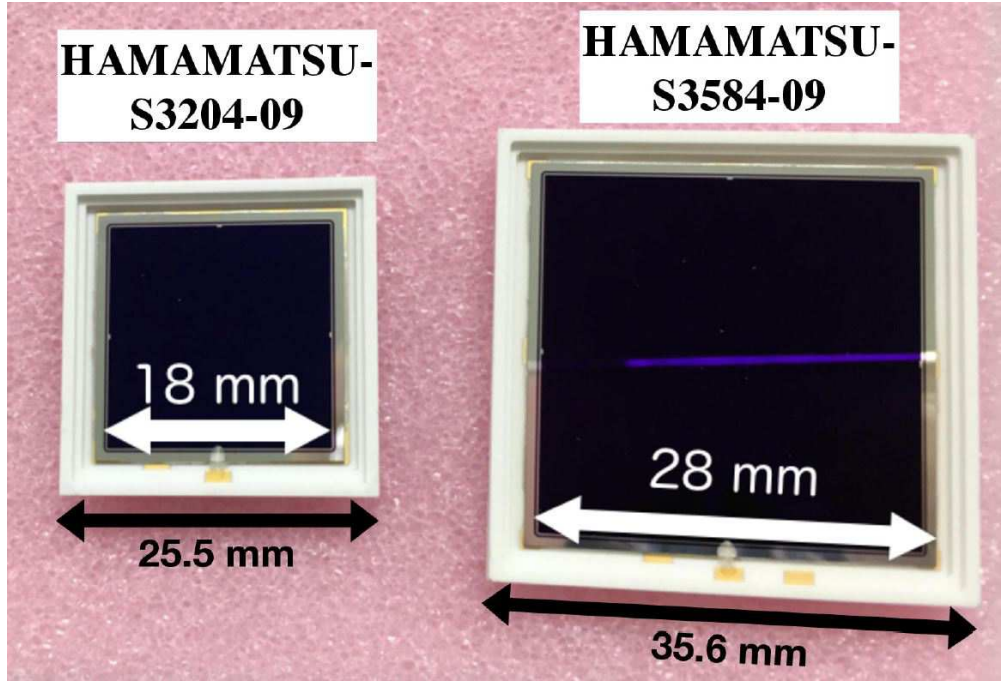}
\caption{A picture of PIN-photodiodes used in this study. The left is HAMAMATSU-S3204-09~($18\times18~\mathrm{mm}$), while the right is HAMAMATSU-S3584-09~($28\times28~\mathrm{mm}$).}
\label{fig_pin}
\end{figure}

In previous studies, we used various sizes of PIN-photodiodes produced by Hamamatsu Photonics K. K., such as S3590-06~($10\times10~\mathrm{mm}$)~\cite{1994125}, S3590-09~($10\times10~\mathrm{mm}$)~\cite{Pronost:2018ghn}, S3204-06~($16\times16~\mathrm{mm}$)~\cite{1997710, TAKEUCHI1999334, Mitsuda:2003eu}, and S3204-09~($18\times18~\mathrm{mm}$)~\cite{Hosokawa:2015koa, Nakano:2017rsy}.  In this study, S3584-09 with a sensitive area of $28\times28~\mathrm{mm}$ was used. Its specifications can be found on the company's website\footnote{\url{https://www.hamamatsu.com/resources/pdf/ssd/s3204-08_etc_kpin1051e.pdf}}. 
Figure~\ref{fig_pin} shows an example of the PIN-photodiodes used in this study. The new PIN-photodiode has approximately $2.5$ times larger surface area than that of previously used PIN-photodiode~(S3204-09, $18\times18~\mathrm{mm}$). In this article, we refer the PIN-photodiode with $18\times18~\mathrm{mm}$ surface area to the small PIN-photodiode while that with $28\times28~\mathrm{mm}$ to the large PIN-photodiode.

As briefly explained in Sect.~ \ref{sec_intro}, the volume where the electrical field is formed increases when the PIN-photodiode with a larger size is installed. Before the calibration described in Sect.~\ref{sec_setup}, we simulated the volume where the electrical field is formed with COSMOL Multiphysics$^{\textcircled{\footnotesize{R}}}$. In this simulation, we assumed a current detector structure with a PIN-photodiode at the top. To quantitatively evaluate the difference, we define the proportion of the volume where the electrical field is formed above $-1.0$~V to the total inner volume of the $80$~L Rn detector as $P_{\mathrm{small}}~(P_{\mathrm{large}})$ when the $18\times18~\mathrm{mm}$~($28\times28~\mathrm{mm}$) PIN-photodiode is installed on the Rn detector. 

\begin{figure}
\centering\includegraphics[width=\linewidth]{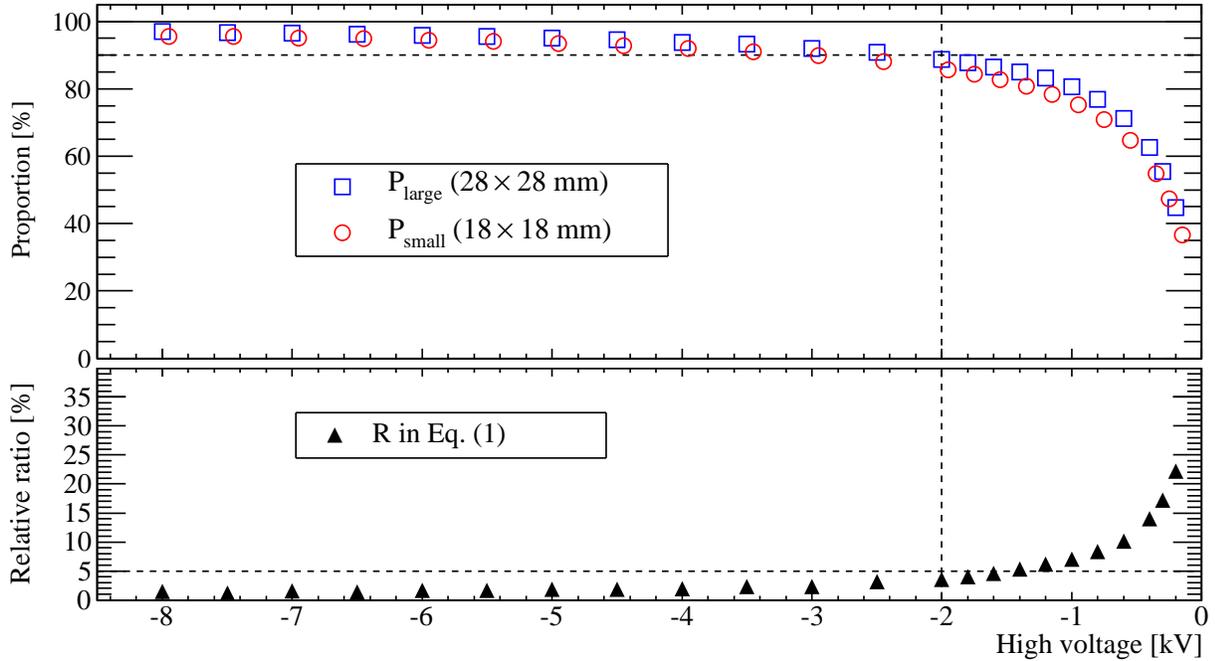}
\caption{Top: The proportion~($P_{\mathrm{small}}$ or $P_{\mathrm{large}}$) of the volume where the electrical field is formed above $-1.0$~V to the total inner volume of the $80$~L Rn detector. The vertical(horizontal) axis shows the ratio in units of percent~(supplied high voltage in units of kV). The blue square~(red circle) points show the simulated result when the PIN-photodiode of $28\times28~\mathrm{mm}$~($18\times18~\mathrm{mm}$) is installed on the $80$~L Rn detector. For ease of visualisation, the red circle shifts horizontally by $-0.25$~kV from the original position. Bottom: The black triangle shows the relative ratio~($R$) defined in Eq.~(\ref{eq_1}).}
\label{fig_active}
\end{figure}

Figure~\ref{fig_active} shows $P_{\mathrm{small}}$ and $P_{\mathrm{large}}$ as functions of the supplied high voltage and their relative ratio, which is defined as
\begin{equation}
R = \frac{P_{\mathrm{large}}-P_{\mathrm{small}}}{P_{\mathrm{small}}} \times 100.0~[\%]. \label{eq_1}
\end{equation}
The volume where the electrical field is formed proportionally increases to the supplied high voltage. The relative ratio~($R$) rapidly increases until the supplied high voltage reaches $-2.0$~kV, while it is almost saturated above this level. For example, $P_{\mathrm{small}}$ is $85.63\%$, whereas  $P_{\mathrm{large}}$ is $88.71\%$ at $-2.0$~kV. The difference between them is $+3.59\%$ at $-2.0$~kV, which demonstrates that the electrical field formed inside the detector is enlarged by replacing the PIN-photodiodes. Therefore, the large PIN-photodiode allows the Rn detector to improve its sensitivity. Below $-2.0$~kV, the difference between $P_{\mathrm{samll}}$ and $P_{\mathrm{large}}$ is significantly large, which implies that the small PIN-photodiode is screened by the surrounding stainless structure.

\subsection{Calibration factor}
Regarding the collection efficiency of the Rn detector, we defined a calibration factor~($C_{F}$) as a parameter to convert the count rate of $\mathrm{^{214}Po}$ in units of count/day into the Rn concentration in units
of $\mathrm{mBq/m^{3}}$. In the calibration described in the next section, we evaluated $C_{F}$ by measuring the count rate of $\mathrm{^{214}Po}$ signals from a particular Rn detector with a constant Rn concentration under radioactive equilibrium in a calibration setup.

\section{Calibration with purified air} \label{sec_setup}
\subsection{Calibration setup}
\begin{figure*}
\centering\includegraphics[width=\linewidth]{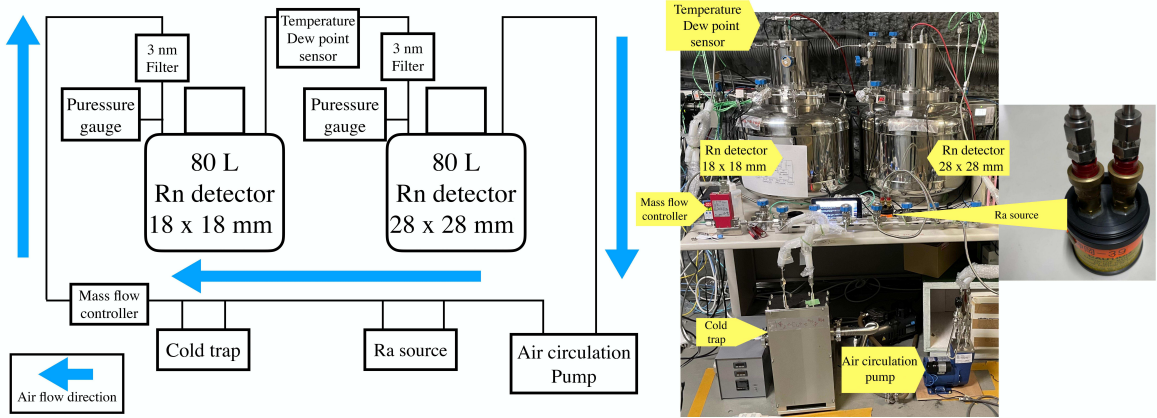}
\caption{Schematic view of the calibration setup~(left) and image of the actual setup~(right). Arrows in the left schematic view show the direction of air circulation.}
\label{fig_setup}
\end{figure*}

We constructed a calibration setup at Lab-A in the experimental area at Kamioka Observatory, the Institute for Cosmic Ray Research~(ICRR), and the University of Tokyo. Figure~\ref{fig_setup} shows the schematic view of the calibration setup and the image of the actual setup.

The system consists of two Rn detectors, a temperature sensor~(VISALA DMT 340), an air circulation pump~(IWAKI BA-330SN), a cold trap~(Taisho TC0147), and pressure gauges~(Naganokeiki Co. Ltd. ZT67), and a mass flow controller~(HORIBA SEC-Z512MGX). We installed a large PIN-photodiode to one of the two Rn detectors and a small PIN-photodiode to the other. To prevent dust from entering the Rn detector, a $3$~nm filter~(Pall Corp. GLFF4281VMM4) was installed in front of each Rn detector. To supply a high voltage to the Rn detectors in parallel, a high-voltage divider~(HAYASHI-REPIC Co., Ltd., RPH-033) was used. To reduce the electrical noise, a noise cut trans~(DENKENSEIKI Research Institute Co., Ltd., NCT-I3) was connected.
 
The temperature sensor monitors the temperature and dew point of the circulating gas. The pressure gauge was also equipped with a Rn detector to monitor the inner pressure. To control the humidity in the gas, we used a cold trap, which is a 1/2-inch U-shaped pipe filled with $14.0$~g of $\phi~80~\mu\mathrm{m}$ copper wool~(Nippon Steel Wool Co. Ltd.). This U-shaped pipe was placed in a heat-insulated vacuum container and connected to a refrigerator.

Other system components are made of electro-polished stainless steel~(NISSHO Astec Co. Ltd. MGS-EP SUS316L), and all joints are connected by VCR$^{\textcircled{\footnotesize{R}}}$ gaskets to minimise Rn emanation and possible air leaks, both of which may affect the background measurements. The air leak rate of the system was measured to be less than $10^{-10}~\mathrm{Pa\cdot m^{3}/s}$ using a helium leak detector~(ULVAC Equipment Sales Inc. HELIOT 712D2). 

To inject Rn gas into the calibration system, we used an Ra source~(PYLON RNC) whose activity is ($199\pm4\%$)~Bq. This calibration source was connected to the system with an adapter of a steel quick connect stem produced by Swagelok, as shown in Figure~\ref{fig_setup}.

\subsection{Calibration procedure}
Calibrations were performed by the following procedures. Before the calibration, we evacuated the entire calibration setup down to $1.0\times10^{-4}$~Pa by a vacuum pump~(Hakuto Co. Ltd. HiCube 80 Eco). Then, we filled the entire calibration system with commercially available G1-grade purified air, whose impurities were less than $0.1$~ppm, until the inner pressure reached the atmospheric pressure. We attached the Ra source to the calibration system and started to circulate the air to maintain a constant Rn concentration in the calibration system. After reaching radioactive equilibrium in the calibration system, we measured the count rate of the $\mathrm{^{214}Po}$ signal from the Rn detector. 

The Rn concentration under radioactive equilibrium was stable at approximately $1243.8\pm49.7~\mathrm{Bq/m^{3}}$, where the total volume of the calibration setup is $0.160~\mathrm{m^{3}}$. Note that the total volume, except for the two Rn detectors, is only $0.6\times10^{-3}~\mathrm{m^{3}}$, and this volume can be ignored in the calculation because the uncertainty of the Rn detector volume is approximately $2$--$3\times10^{-3}~\mathrm{m^{3}}$ level, as mentioned in the later section~(Sect.~\ref{sec_sys}). Finally, we evaluated the calibration factors~($C_{F}$) by comparing the count rate and constant Rn concentration.

\subsection{High-voltage dependence}
\begin{figure*}
\begin{minipage}{0.5\hsize}
\centering\includegraphics[width=1.0\linewidth]{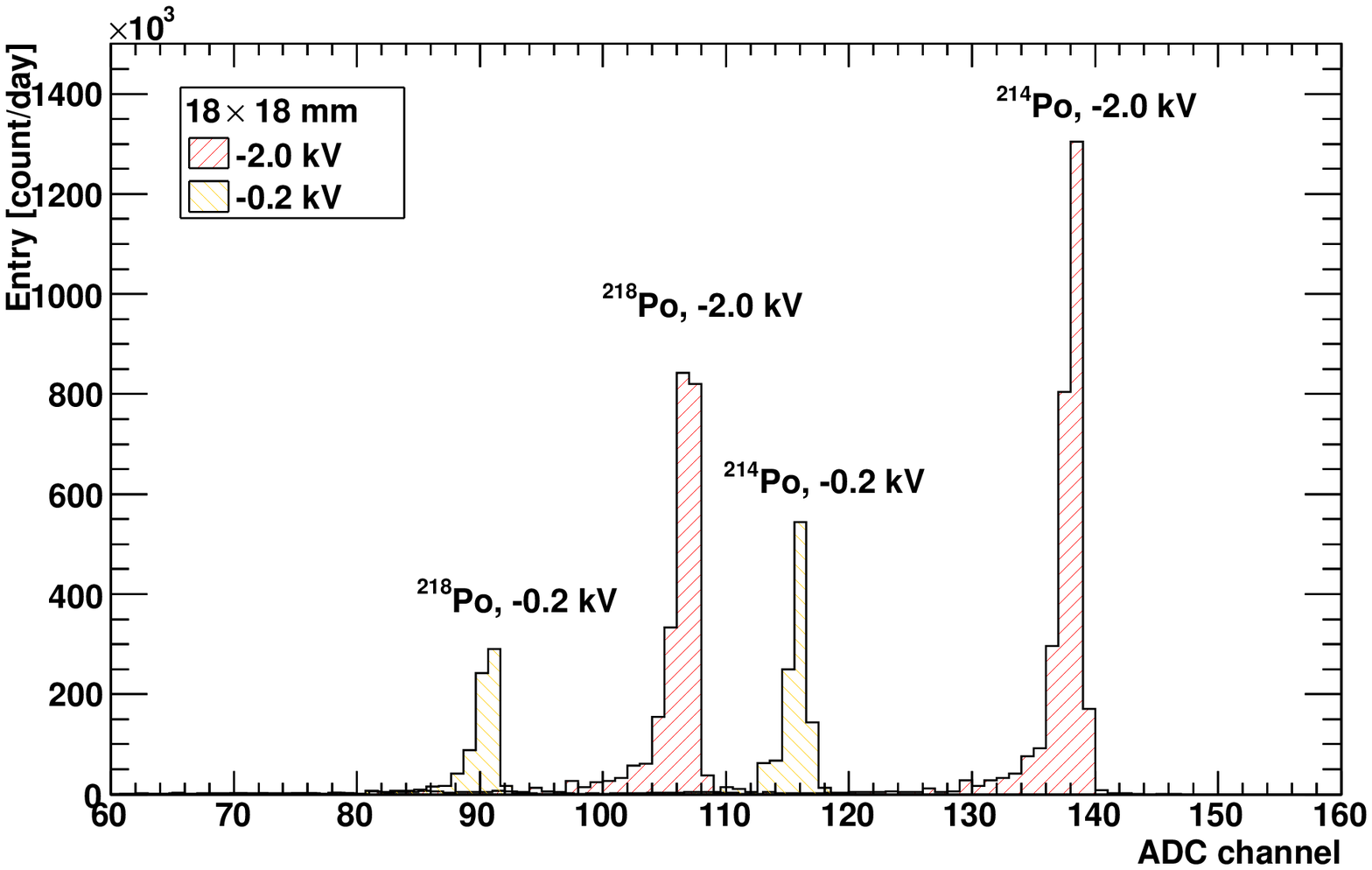}
\end{minipage}
\begin{minipage}{0.5\hsize}
\centering\includegraphics[width=1.0\linewidth]{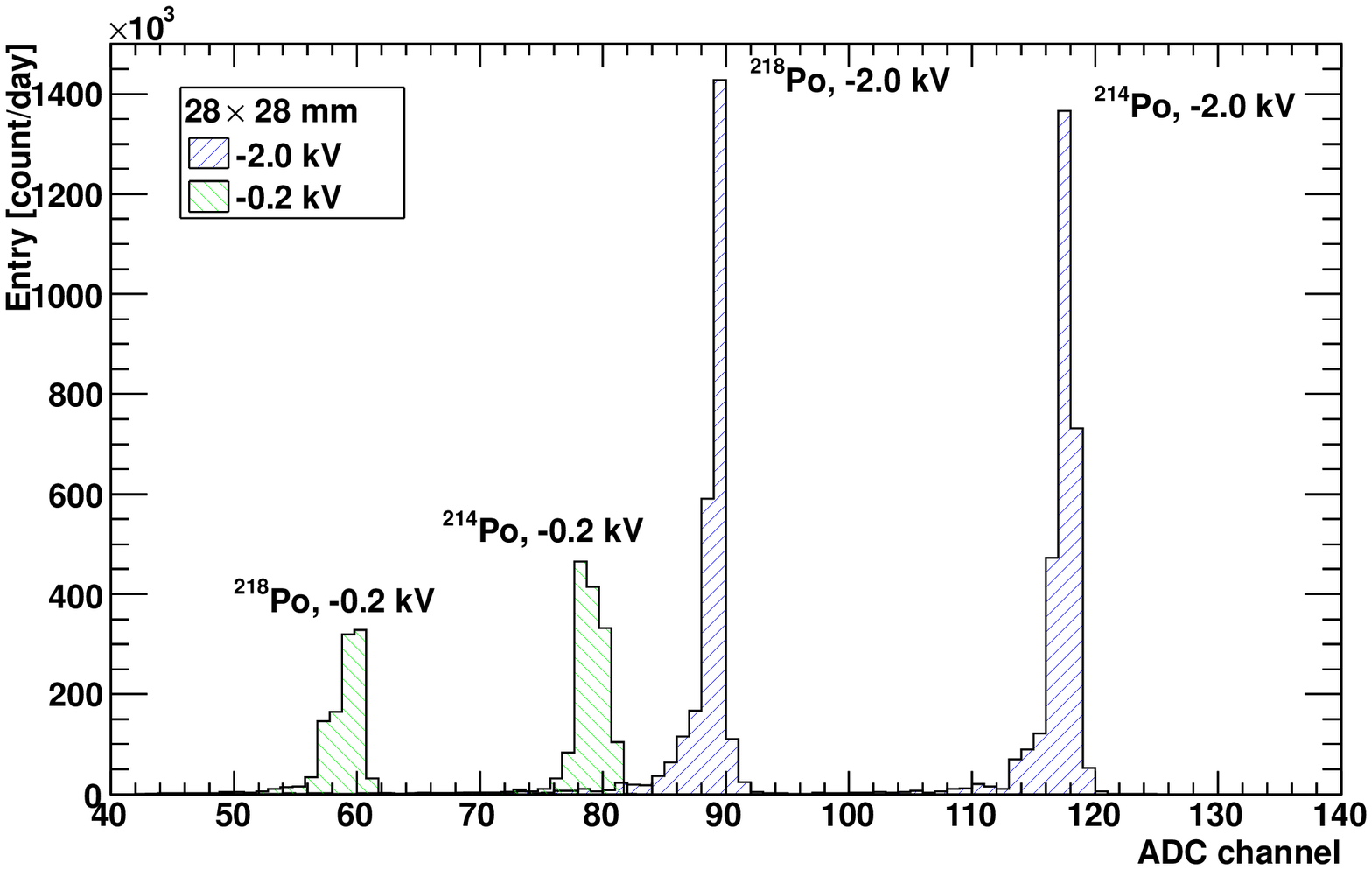}
\end{minipage}
\caption{Typical daily spectrum measured by the Rn detector equipped with  $18\times18~\mathrm{mm}$~($28\times28~\mathrm{mm}$) PIN-photodiode in left (right). As an example, the spectrum shown here was measured under radioactive equilibrium by supplying high voltage at $-2.0$~kV~(red in left and blue in right) and $-0.2$~kV~(yellow in left and green in right).}
\label{fig_spect_shift}
\end{figure*}

Because the electrical field between the PIN-photodiode and the stainless steel vessel is formed by the supplied high voltage, the collection efficiency is highly dependent on the supplied high-voltage value. To confirm this dependence, we evaluated the calibration factor by changing the supplied high voltage. Figure~\ref{fig_spect_shift} shows examples of the daily spectrum measured by the Rn detector equipped with the small~(large) PIN-photodiode during the calibration.

There are two large peaks in the spectrum originating from $\mathrm{^{218}Po~(6.00~MeV)}$ and $\mathrm{^{214}Po~(7.69~MeV)}$. The $\mathrm{^{214}Po}$ signal is used to calculate the Rn concentration because there are no other $\alpha$~sources near its peak energy\footnote{In the case of the $\mathrm{^{218}Po}$ signal, several background sources from the $\mathrm{^{232}Th}$ series exist. In particular, $\alpha$ particles are emitted from $\mathrm{^{208}Tl}~(5.00~\mathrm{MeV})$,  $\mathrm{^{228}Th}~(5.52~\mathrm{MeV})$, $\mathrm{^{224}Ra}~(5.79~\mathrm{MeV})$, $\mathrm{^{212}Bi}~(6.21~\mathrm{MeV})$, and $\mathrm{^{220}Rn}~(6.41~\mathrm{MeV})$. Their signals potentially overlapped with the peak energy of $\mathrm{^{218}Po}$. Thus, $\mathrm{^{218}Po}$ signals are not used to calculate the Rn concentration in the gases.}. Even though the PIN-photodiode has a high spectral resolution, background signals from $\mathrm{^{216}Po~(6.78~MeV)}$ and $\mathrm{^{212}Po~(8.95~MeV)}$ potentially overlap with the signals from $\mathrm{^{214}Po}$. Hence, the signal window must be properly determined to evaluate the calibration factor. For this reason, we set the signal window to count $\mathrm{^{214}Po}$ signals based on the criteria in a previous publication~\cite{Nakano:2017rsy}. We set the lower bound at five channels from the peak of $\mathrm{^{216}Po}$ signals not only to prevent the entry of the $\mathrm{^{216}Po}$ signals but also to cover the tail of the $\mathrm{^{214}Po}$ signals. On the other hand, we set the upper bound at five channels from the peak of $\mathrm{^{214}Po}$ signals to prevent the entry of $\mathrm{^{212}Po}$ signals. Note that $\mathrm{^{216}Po}$ and $\mathrm{^{212}Po}$ originate from the decay of $\mathrm{^{220}Rn}$, and their observed rates are quite low compared with that of $\mathrm{^{214}Po}$ signals during the calibration.

The pulse height of the output signal shifts when the supplied high-voltage changes because the electronics simultaneously supply the reverse bias minus the high voltage to the PIN-photodiode. Figure~\ref{fig_spect_shift} also shows the daily spectrum by changing the supplied high voltage from $-0.2$~kV to $-2.0$~kV. As expected, the peak positions of the observed $\mathrm{^{214}Po}$~(and $\mathrm{^{218}Po}$) signals shift depending on the supplied high voltage. Furthermore, the total number of daily observed signal rates decreases when the supplied high voltage is lowered.

\begin{figure}
\centering\includegraphics[width=\linewidth]{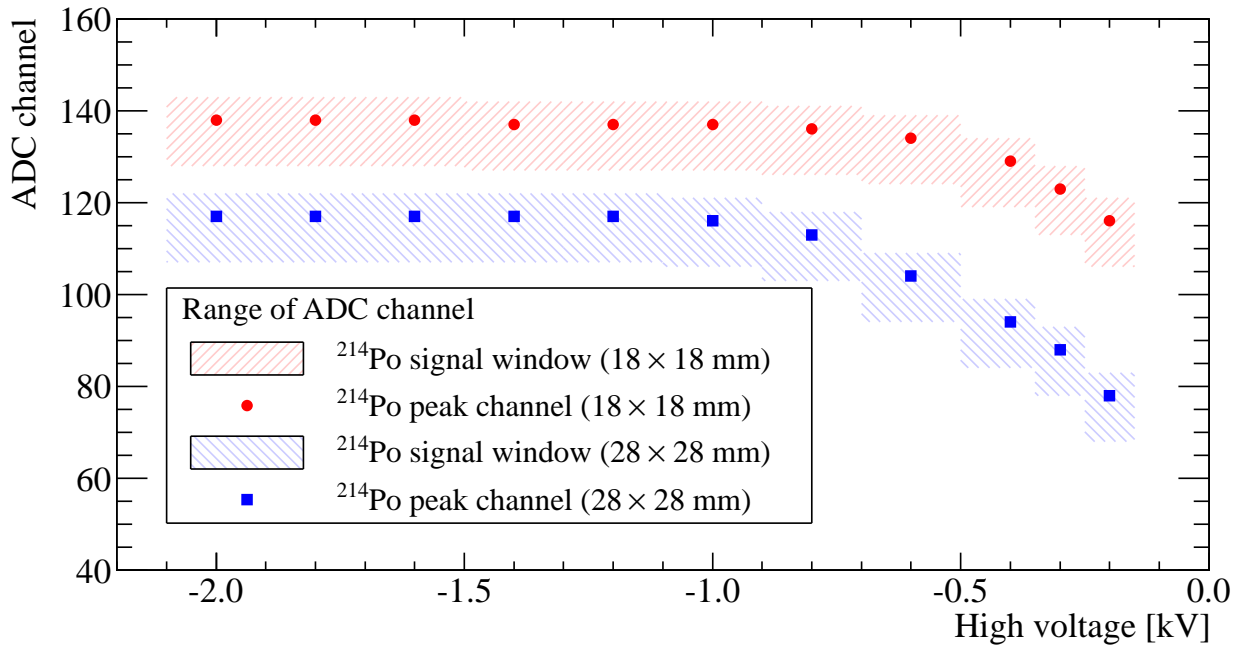}
\caption{The region of the ADC channel determined from the peak position of $\mathrm{^{214}Po}$ signals observed by the PIN-photodiode as a function of the supplied high voltage~[kV]. The red circle~(blue square) points show the ADC channel of the signal peak position for the PIN-photodiode of $18\times18~\mathrm{mm}$~($28\times28~\mathrm{mm}$). The red~(blue) bands show the windows for selecting $\mathrm{^{214}Po}$ signals based on the criteria used in a previous study~\cite{Nakano:2017rsy}.}
\label{fig_range}
\end{figure}

Figure~\ref{fig_range} shows the selected signal windows for the calibration as a function of the supplied high voltage. The peak position of $\mathrm{^{214}Po}$ signals relatively shifts significantly below $-1.0$~kV while the peak position does not shift considerably above $-1.0$~kV.

\begin{figure}
\centering\includegraphics[width=\linewidth]{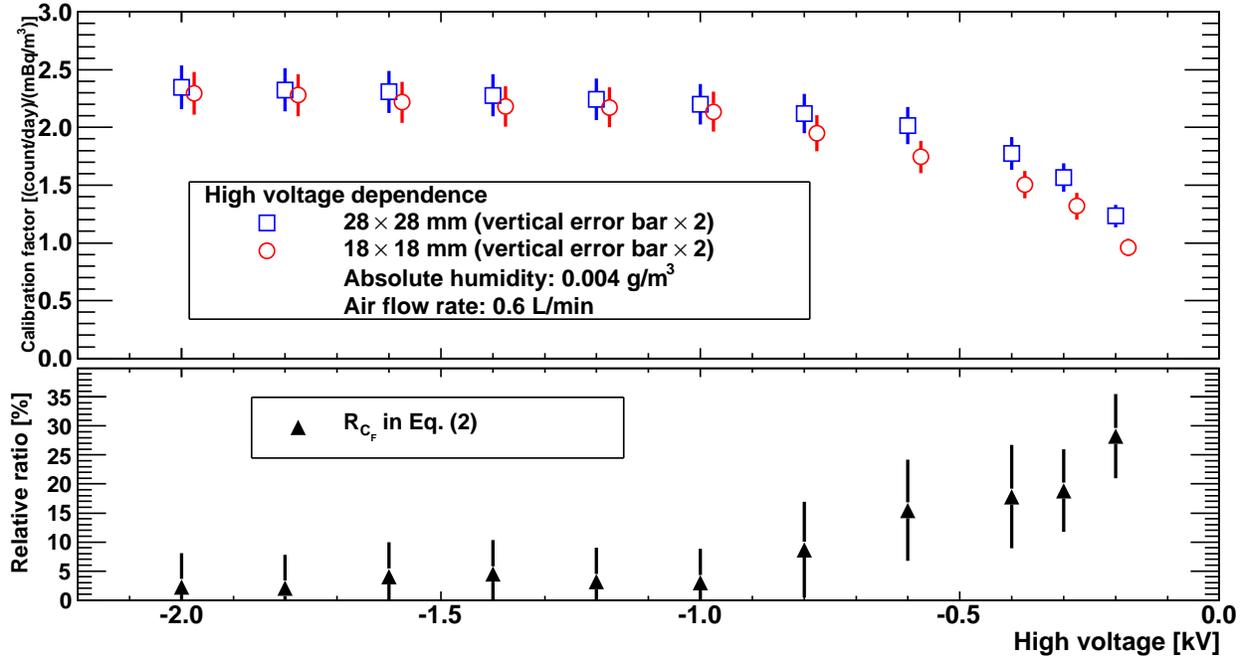}
\caption{The supplied high-voltage dependence of the calibration factor. In the top panel, the blue square~(red circle) points show the calibration factor of the Rn detector equipped with a $28\times28~\mathrm{mm}$~($18\times18~\mathrm{mm}$) PIN-photodiode determined by changing the supplied high voltage. The calibrations were performed by setting the air circulation rate to $0.6$L/min. To ease visualisation, the vertical error bars were scaled by $2$, and the red circles horizontally shifted by $-0.25$~kV from the original position. The absolute humidity during this calibration was controlled at $0.004~\mathrm{g/m^{3}}$. In the bottom panel, the black triangle points show their relative ratio~($R_{C_{F}}$) defined in Eq.~(\ref{eq_2}).}
\label{fig_hv_dep}
\end{figure}

By using these signal windows, we obtained the calibration factors for each high-voltage value. Figure~\ref{fig_hv_dep} shows the calibration factors as functions of the supplied high voltage. To quantitatively evaluate the difference between the two PIN-photodiodes, the relative ratio between their calibration factors, which is defined as
\begin{equation}
    R_{C_{F}} = \frac{C_{F,\,\mathrm{large}}-C_{F,\,\mathrm{small}}}{C_{F,\,\mathrm{small}}} \times 100.0~[\%], \label{eq_2}
\end{equation}

\noindent 
is also calculated for each high-voltage setting. Based on the measurements, we found a $15$--$30\%$-level of improvement below $-1.0$~kV, while only a few~$\%$ level of change occurred above this level. This behaviour is similar to that observed in the region where the electrical field is formed, as shown in Figure~\ref{fig_active}. 

We chose $-2.0$~kV for the calibration of absolute humidity as well as the background rate measurement in the later sections because the divided high-voltage supply to the PIN-photodiode becomes larger than its tolerance limit. Note that the calibration factor is almost saturated above $-2.0$~kV based on the simulation shown in Figure~\ref{fig_active}, and this setting does not significantly affect the calibration result.

\subsection{Absolute humidity dependence} \label{sec_ah}
The collection efficiency also depends on the humidity of the gas because the charged nuclei are captured by water. Their neutralisation results in a loss of collection efficiency~\cite{osti_5305900, howard1991}. To test this dependence, we calibrated $C_{F}$ by changing the temperature of the cold trap from $-80\mathrm{^{\circ}C}$ to room temperature~(typically $+20\mathrm{^{\circ}C}$ in the Kamioka mine). 

\begin{figure}
\centering\includegraphics[width=\linewidth]{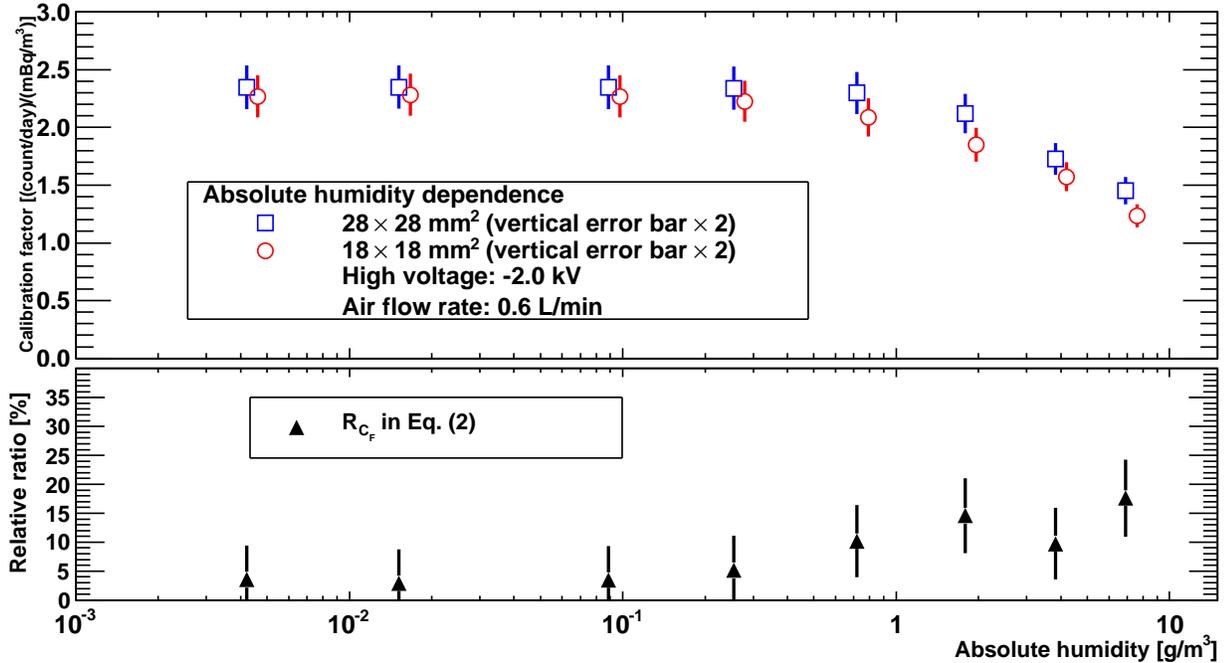}
\caption{The absolute humidity dependence of the calibration factor. In the top panel, the blue square~(red circle) points show the calibration factor determined by the Rn detector equipped with a $28\times28~\mathrm{mm}$~($18\times18~\mathrm{mm}$) PIN-photodiode. The calibrations were performed by setting a high voltage of $-2.0$~kV and an air circulation rate of $0.6$~L/min. To ease visualisation, the vertical error bars are scaled by $2$, and the red circles horizontally shift by $-10\%$ from the original position. In the bottom panel, the black triangle points show the relative ratio~($R_{C_{F}}$) defined in Eq.~(\ref{eq_2}).}
\label{fig_ah_dep}
\end{figure}

Figure~\ref{fig_ah_dep} shows the calibration factors as a function of the absolute humidity of filling purified air and the relative ratio between them~($R_{C_{F}}$). By comparing the calibration factor of the small PIN-photodiode with the large one, we found a level of improvement of $10$--$20\%$ above $1.0~\mathrm{g/m^{3}}$ while a $(3.8\pm2.4)\%$-improvement was observed below this level by replacing the PIN-photodiode.

\begin{figure}
\centering\includegraphics[width=\linewidth]{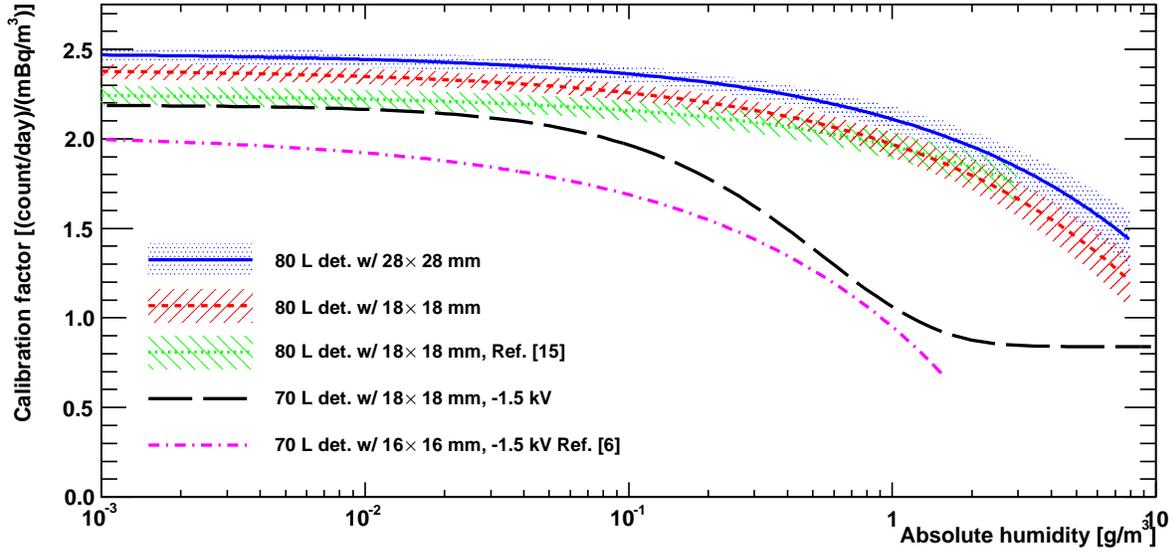}
\caption{Comparison of the absolute humidity dependence of the calibration factor among the Rn detectors when the purified air is filled in. The blue and red lines~(with band) show the empirical functions obtained in this study. The green, black, and pink lines show the results of previous studies~\cite{TAKEUCHI1999334, Nakano:2017rsy}.}
\label{fig_ah_band}
\end{figure}

The behaviour of this dependence is well described by an empirical function, $C_{F}=A-B\sqrt{A_{H}}$, where $A$ and $B$ are parameters to be fitted, and $A_{H}$ is the absolute humidity of the filling gas~\cite{osti_6605848}. Figure~\ref{fig_ah_band} shows a comparison of the empirical function obtained in this study as well as those of previous studies~\cite{TAKEUCHI1999334, Nakano:2017rsy}, where the fitting parameters are summarized in Appendix~\ref{sec_app1}.
 
In the case of the $70$~L Rn detector, the acrylic plate and Viton O-rings were used as the top flanges of the detector. Because of this design, the detector is not airtight, and therefore its sensitivity is worse in humid conditions~\cite{TAKEUCHI1999334}. After replacing the top acrylic flange with knife-edge flanges with metal gaskets, the calibration factors in humid conditions were improved~\cite{Hosokawa:2015koa, Nakano:2017rsy}.

By comparing the shapes of the empirical functions, we found that the $80$~L Rn detector with the PIN-photodiode of $28\times28~\mathrm{mm}$ has the best sensitivity among them. This improvement under dry conditions below $0.1~\mathrm{g/m^{3}}$ results from replacing the PIN-photodiode.

\subsection{Systematic uncertainty in the calibration} \label{sec_sys}
The systematic uncertainties are estimated according to the technical specifications of the measurement devices. The assigned systematic uncertainties are summarized in Table~\ref{table_sys}.

\begin{table}
\caption{Summary of the systematic uncertainties in the calibration.}
\label{table_sys}
\centering
\begin{tabular}{cc}
\hline
Item & Assigned \\ 
 & uncertainty \\
\hline
Ra source activity& $\pm4\%$\\
Temperature sensor& $\pm2\%$\\
Air flow rate& $\pm2\%$\\
Volume of the whole calibration setup& $\pm3\%$\\
\hline
\end{tabular}
\end{table}

Based on the manufacturer's documentation, the uncertainty of the activity of the Ra source is $\pm4\%$. The temperature sensor has an uncertainty of $\pm2\%$, and the mass flow controller has an uncertainty of $\pm2\%$. Because the volume of the whole calibration system is not measured precisely, we assigned the systematic error of the total volume as $\pm3\%$, where the dominant uncertainty comes from the accuracy of the Rn detector's structure. These systematic uncertainties are included in the calibration results described in this section.

\section{Background measurement} \label{sec_bg}

\subsection{Background rate with purified air}
The intrinsic background rate was evaluated by closing the detector. This measurement evaluates the intrinsic background due to the emanation from the inner surface of the Rn detector, including the PIN-photodiode, because some amount of Rn emanating from other components of the calibration setup increases the background rate. To eliminate such uncertainties, we measured the background rate by closing the Rn detectors. In addition, the background from the calibration setup was also evaluated by circulating air. The result is described in Appendix~\ref{sec_bg_setup}.

Figure~\ref{fig_bg_spect} shows the measured daily background spectrum of the $80$~L Rn detector with $28\times28~\mathrm{mm^{2}}$ PIN-photodiode\footnote{The signal region is slightly different from that used in Figure~\ref{fig_range} since we replaced the electronics after the calibration described in Sect.~\ref{sec_setup}.} and the average background count rate\footnote{Once the Rn detector is closed, the residual Rn in the purified air starts to decay, and Rn also starts to emanate from the detector's inner surface. They contribute as backgrounds. To consider these two physics processes, it is proper to fit the background rate with the following function:
\begin{equation}
  Ae^{-\lambda t} + B(1-e^{-\lambda t}), \notag
\end{equation}
\noindent
where $t$ is the elapsed time since the start of the measurement, $\lambda$ is the decay constant for $\mathrm{^{222}Rn}$, and $A$ and $B$ are parameters to be fitted. However, the background rate shown in Figure~\ref{fig_bg_spect} is quite low. Hence, we calculated the average count rate as the background for this study.}. The count rate is $0.57^{+0.20}_{-0.10}$~count/day, which corresponds to a Rn concentration of $0.24^{+0.09}_{-0.05}~\mathrm{mBq/m^{3}}$. Table~\ref{table_bg} summarizes the background rates evaluated in this study, as well as those measured in previous studies. The intrinsic background of the $80$~L Rn detector with $28\times28~\mathrm{mm}$ is consistent with that of the $18\times18~\mathrm{mm}$ evaluated in previous studies~\cite{Hosokawa:2015koa, Nakano:2017rsy}. Moreover, their background rates were unchanged, although we installed a larger PIN-photodiode.

\begin{figure*}
\begin{minipage}{0.5\hsize}
\centering\includegraphics[width=1.0\linewidth]{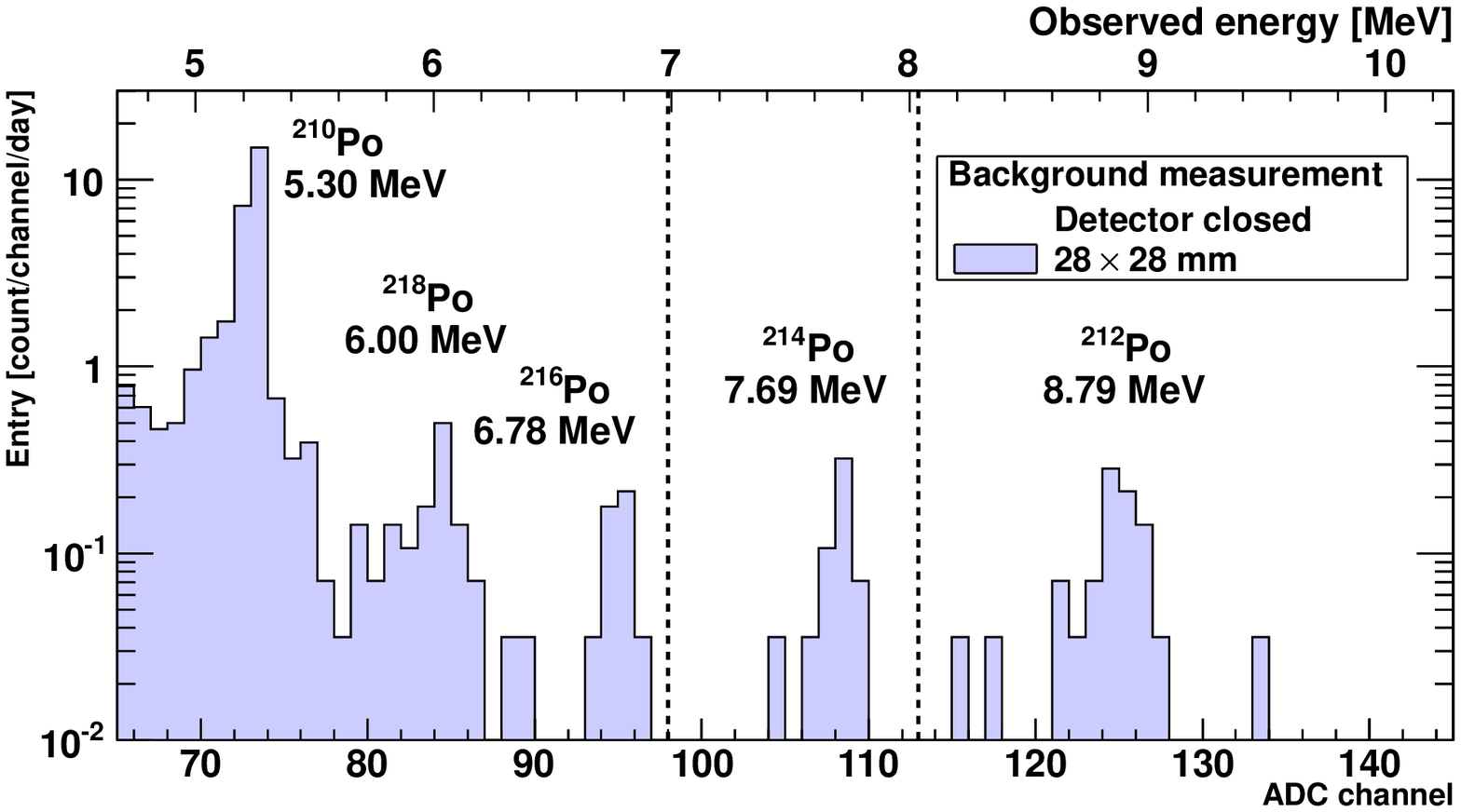}
\end{minipage}
\begin{minipage}{0.5\hsize}
\centering\includegraphics[width=1.0\linewidth]{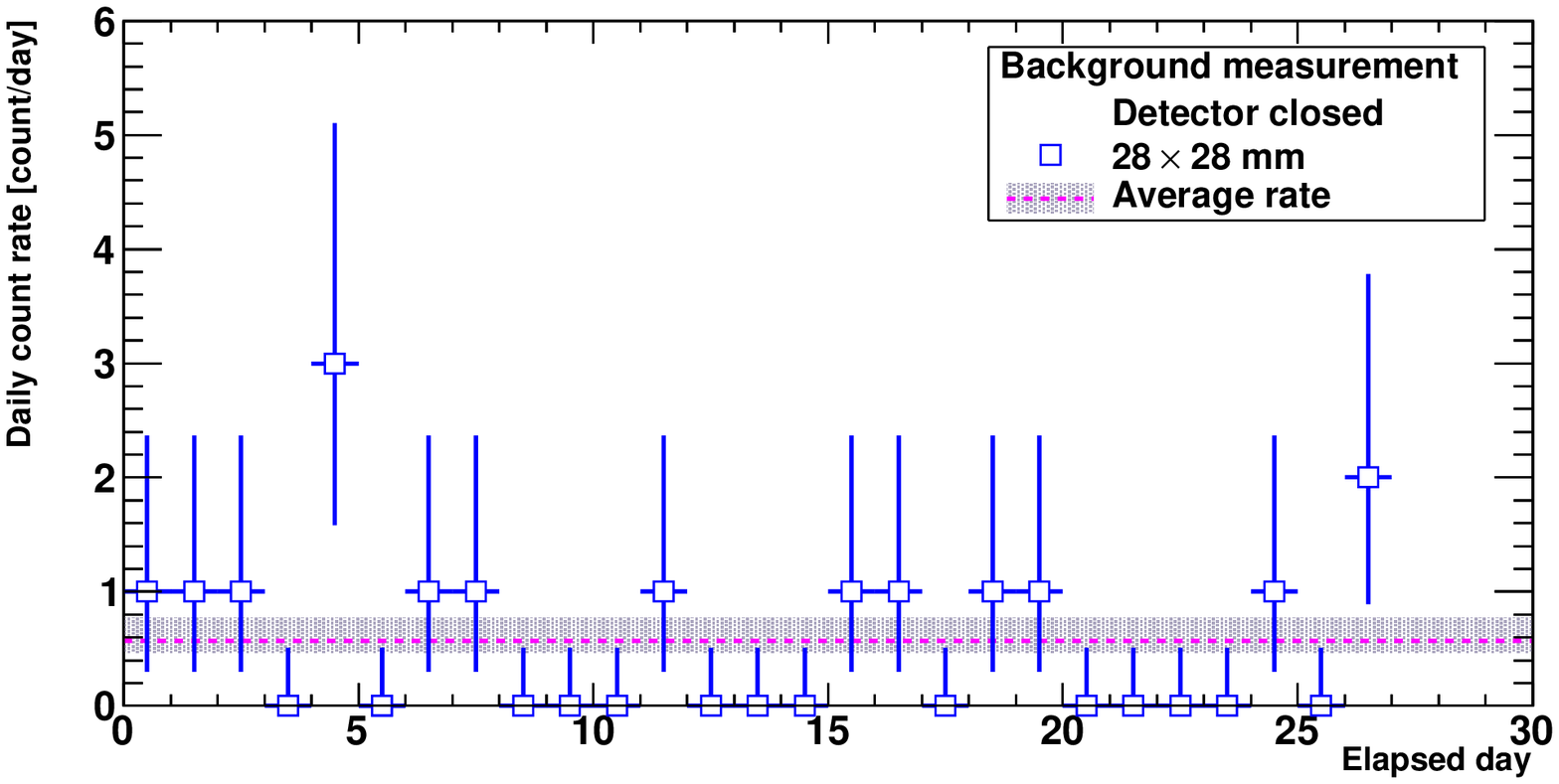}
\end{minipage}
\caption{(Left) The measured background spectrum of the $80$~L Rn detector with $28\times28~\mathrm{mm}$ of PIN-photodiode. The vertical axis shows the count/channel/day, and the bottom~(top) horizontal axis shows the ADC channel~(measured energy in units of MeV). The two dashed lines show the upper and lower bounds of the $\mathrm{^{214}Po}$ signal region. (Right) The $\mathrm{^{214}Po}$ count rate of the 80~L Rn detector with $28\times28~\mathrm{mm}$ of the PIN-photodiode is a function of time. The blue square points show the daily count rate, and the dashed horizontal line~(gray band) shows the average~(uncertainty) of the count rate.}
\label{fig_bg_spect}
\end{figure*}

\begin{table*}
\centering
\caption{Summary of the intrinsic background rates of the $80$~L Rn detector equipped with the $28\times28~\mathrm{mm}$~($18\times18~\mathrm{mm}$) PIN-photodiode. These measurements were conducted by filling the purified air, and the absolute humidity was controlled to be less than $10^{-3}~\mathrm{g/m^{3}}$.}
\label{table_bg}
\begin{tabular}{cccc}
\hline
 & Livetime & Rn detector & Rn detector  \\ 
 & [day] &  with $28\times28~\mathrm{mm}$ & with $18\times18~\mathrm{mm}$ \\
 \hline
 This study & $28$ & $0.57^{+0.20}_{-0.10}$~count/day & --  \\
 ($80$~L Rn detector) & & $0.24^{+0.09}_{-0.05}~\mathrm{mBq/m^{3}}$ & --   \\ 
\hline \hline
 Past study 1~\cite{TAKEUCHI1999334} & Approximately  &  -- &  $2.4\pm1.3$~count/day \\ 
 ($70$~L Rn detector) & 2~months & -- & -- \\ 
\hline
 Past study 2~\cite{Hosokawa:2015koa} & $132.5$ &  -- &  $0.81\pm0.08$~count/day \\ 
 ($80$~L Rn detector) & & -- & $0.37\pm0.05~\mathrm{mBq/m^{3}}$ \\ 
\hline
 Past study 3~\cite{Nakano:2017rsy} & $156$ &  -- &  $0.78\pm0.07$~count/day \\ 
 ($80$~L Rn detector) & & -- & $0.33\pm0.07~\mathrm{mBq/m^{3}}$ \\ 
\hline
\end{tabular}
\end{table*}

Relatively large signals from $\mathrm{^{210}Po}$ are observed, as shown in Figure~\ref{fig_bg_spect}. These signals come from the $\mathrm{^{222}Rn}$ daughters remain on the PIN-photodiode once they have been collected because the intrinsic background measurement was conducted after the calibrations described in Sect.~\ref{sec_setup}.

The daily Rn emanation rate from the $80$~L Rn detector can be estimated by considering the correction of the decay duration since its generation. The correction factor, defined as $\varepsilon$, is expressed as 
\begin{align}
\varepsilon & = \frac{\mathrm{Number~of~Rn~after~emanation}}{\mathrm{Number~of~Rn~under~radioactive~equilibrium}} \notag \\ 
 & = \frac{N\left\{1-\left(\frac{1}{2}\right)^{1.0/3.82}\right\} }{N\left\{1-\left(\frac{1}{2}\right)^{\infty/3.82}\right\} } = 0.1659. \label{eq_3}
\end{align}

\noindent Using this correction factor, the daily emanation rate from the $80$~L Rn detector is calculated as $(3.18^{+1.19}_{-0.66})~\mathrm{\mu Bq/detector/day}$.

\section{Application of the Rn detector with a large PIN-photodiode} \label{sec_future}
\subsection{Rn measurement of purified water}

As prospects, upgrading by replacing the larger PIN-photodiode is helpful to precisely monitor its Rn concentration in a humid gas, as described in Sect.~\ref{sec_ah}. In the case of Super-Kamiokande, the Rn concentration in the purified water was measured by the Rn detector combined with a membrane degasifier module~\cite{TAKEUCHI1999334, Mitsuda:2003eu}. However, the sensitivity of this method is limited by the inefficiency of the Rn detector in a humid gas after extracting Rn gas from the purified water. Another approach using both an extraction column and a cooled charcoal trap is adapted at the expense of real-time measurement instead of the membrane degasifier module~\cite{Nakano:2019bnr}. By replacing the large PIN-photodiode and the newly developed membrane degasifier module with a low intrinsic background~\cite{takeda:nu2020}, the sensitivity of the Rn detector combined with the membrane degasifier module is expected to be improved.

\subsection{Material screening by measuring the Rn emanation rate}

In rare event search experiments, the radioactive impurities in the detector's material are severe background sources. The emanation rate is fixed once the detector is constructed. For this reason, careful material screening programs with several screening devices have been conducted before the construction of each detector ~\cite{Leonard:2007uv, Aprile:2011ru, Wang:2016eud, Aprile:2017ilq, Cebrian:2017jzb}.

The Rn detector is used to measure the Rn emanation rate from materials by combining it with a sampling chamber in which the material is put in~\cite{Sekiya:2015rla,Takeuchi:2017pij}. In such a measurement system, water is also emitted from the surface of the material, resulting in the neutralisation of Rn daughters. Hence, the control of absolute humidity is the key to precisely evaluating the material's Rn emanation rate. In this study, we found a higher collection efficiency in humid gases by replacing a small PIN-photodiode with a larger one. Therefore, this upgrade also helps to apply the Rn detector to such screening programs to evaluate the impurities in the detector's materials.

\section{Summary and conclusion} \label{sec_conclusion}
To improve the detection sensitivity for measuring Rn in gases, we developed a new Rn detector by replacing the PIN photodiode, HAMAMATSU-S3584-09. To evaluate the detector performance, we constructed a calibration setup by serially connecting two Rn detectors equipped with both types of PIN photodiodes. 

First, we tested the high-voltage dependence of the calibration factor. Based on the calibrations, we found a $15$--$30\%$-level of improvement of the calibration factor below $-0.3$~kV, while a few~$\%$ improvements occurred above this level. We also tested the absolute humidity dependence of the calibration factor. We found an improvement of $(3.8\pm2.4)\%$ in the detection efficiencies below $1.0~\mathrm{g/m^{3}}$, while a $10$--$20\%$-improvement occurred above this level. We also measured the intrinsic background of the detector, $0.24^{+0.09}_{-0.05}~\mathrm{mBq/m^{3}}$. This background is consistent with the Rn detector of the small PIN-photodiode. We successfully demonstrated that the sensitivity of the detector was improved by replacing the larger PIN-photodiode.

The limitation of the collection efficiency in the current design comes from the difficulty in forming the electrical field between the PIN-photodiode and the stainless steel structure because of its small volume. Further sensitivity is expected by enlarging the Rn detector's volume.

\section*{Acknowledgment}

We gratefully acknowledge the cooperation of the Kamioka Mining and Smelting Company. We thank A.~Takeda from Kamioka Observatory, ICRR, the University of Tokyo, for preparing the additional electrical supply line at the experimental site. K. O. and Y. N. thank T. Onoue for constructing the support structures for the calibration setup. This work was partially supported by MEXT KAKENHI Grant Numbers 18H05536, 19J21344, 20K03998, and 21K13942. This work was partially supported by an inter-university research program at ICRR.


\bibliographystyle{ptephy}
\bibliography{main.bib}

\begin{thebibliography}{10}

\bibitem{4329405}
V.C. Negro and S.~Watnick, IEEE Trans. Nucl. Sci, {\bf 25}, 757--761 (1978).

\bibitem{HOWARD1990589}
A.J. Howard, B.K. Johnson, and W.P. Strange, Nucl. Instrum. Meth. A, {\bf 293},
  589--595 (1990).

\bibitem{1994125}
S.~Tasaka, Y.~Sasaki, H.~Okazawa, and M.~Nakagawa, RADIOISOTOPES, {\bf 43},
  125--133 (1994).

\bibitem{1996741}
S.~Tasaka and Y.~Sasaki, RADIOISOTOPES, {\bf 45}, 741--752 (1996).

\bibitem{1997710}
M.~Nemoto, S.~Tasaka, H.~Hori, K.~Okumura, T.~Kajita, and Y.~Takeuchi,
  RADIOISOTOPES, {\bf 46}, 710--719 (1997).

\bibitem{TAKEUCHI1999334}
Y.~Takeuchi, K.~Okumura, T.~Kajita, S.~Tasaka, H.~Hori, M.~Nemoto, and
  H.~Okazawa, Nucl. Instrum. Meth. A, {\bf 421}, 334--341 (1999).

\bibitem{CHOI2001177}
E.~Choi, M.~Komori, K.~Takahisa, N.~Kudomi, K.~Kume, K.~Hayashi, S.~Yoshida,
  H.~Ohsumi, H.~Ejiri, T.~Kishimoto, K.~Matsuoka, and S.~Tasaka, Nucl. Instrum.
  Meth. A, {\bf 459}, 177--181 (2001).

\bibitem{KIKO2001272}
J.~Kiko, Nucl. Instrum. Meth. A, {\bf 460}, 272--277 (2001).

\bibitem{Mitsuda:2003eu}
C.~Mitsuda, T.~Kajita, K.~Miyano, S.~Moriyama, M.~Nakahata, Y.~Takeuchi, and
  S.~Tasaka, Nucl. Instrum. Meth. A, {\bf 497}, 414--428 (2003).

\bibitem{GUTIERREZ2004583}
J.L. Guti\'{e}rrez, M.~Gar\'{c}a-Talavera, V.~Pe$\tilde{\mathrm{n}}$a, J.C.
  Nalda, M.~Voytchev, and R.~L\'{o}pez, Appl. Radiat. Isot., {\bf 60}, 583--587
  (2004).

\bibitem{MARTINMARTIN20061287}
A.~Martin-Martin, J.L. Gutierrez-Villanueva, J.M. Munoz, M.~Garcia-Talavera,
  G.~Adamiec, and M.P. Iniguez, Appl. Radiat. Isot., {\bf 64}, 1287--1290
  (2006).

\bibitem{Mamedov:2011zz}
F.~Mamedov, I.~Stekl, J.~Konicek, K.~Smolek, and P.~Cermak, JINST, {\bf 6},
  C12008 (2011).

\bibitem{2013InJPh..87..471A}
P.~{Ashokkumar}, B.K. {Sahoo}, A.~{Topkar}, A.~{Raman}, D.A.R. {Babu}, D.N.
  {Sharma}, and Y.S. {Mayya}, Indian J. Phys., {\bf 87}, 471--477 (2013).

\bibitem{Hosokawa:2015koa}
K.~Hosokawa, A.~Murata, Y.~Nakano, Y.~Onishi, H.~Sekiya, Y.~Takeuchi, and
  S.~Tasaka, PTEP, {\bf 2015}, 033H01 (2015).

\bibitem{Nakano:2017rsy}
Y.~Nakano, H.~Sekiya, S.~Tasaka, Y.~Takeuchi, R.A. Wendell, M.~Matsubara, and
  M.~Nakahata, Nucl. Instrum. Meth. A, {\bf 867}, 108--114 (2017),
  {{arXiv:1704.06886}}.

\bibitem{Pronost:2018ghn}
G.~Pronost, M.~Ikeda, T.~Nakamura, H.~Sekiya, and S.~Tasaka, PTEP, {\bf 2018},
  093H01 (2018),  {{arXiv:1807.11142}}.

\bibitem{Elisio:2019nuz}
S.~Elisio and L.~Peralta, Nucl. Instrum. Meth. A, {\bf 969}, 164033 (2020),
  {{arXiv:1907.08143}}.

\bibitem{Kotrappa1981ElectretaNT}
P.~Kotrappa, S.~Dua, P.C. Gupta, and Y.S. Mayya, Health Phys., {\bf 41}, 35--46
  (1981).

\bibitem{PMID:2663781}
P.K. Hopke, Health Phys., {\bf 57}, 39--42 (1989).

\bibitem{Nakano:2019bnr}
Y.~Nakano, T.~Hokama, M.~Matsubara, M.~Miwa, M.~Nakahata, T.~Nakamura,
  H.~Sekiya, Y.~Takeuchi, S.~Tasaka, and R.A. Wendell, Nucl. Instrum. Meth. A,
  {\bf 977}, 164297 (2020),  {{arXiv:1910.03823}}.

\bibitem{Ogawa:2019ccp}
H.~Ogawa, K.~Abe, M.~Matsukura, and H.~Mimura, JINST, {\bf 15}, P01039 (2020),
  {{arXiv:1910.02617}}.

\bibitem{Nakano:2020tej}
Y.~Nakano, K.~Ichimura, H.~Ito, T.~Okada, H.~Sekiya, Y.~Takeuchi, S.~Tasaka,
  and M.~Yamashita, PTEP, {\bf 2020}, 113H01 (2020),  {{arXiv:2003.11705}}.

\bibitem{Abe:2011ts}
K.~Abe et~al.,
\newblock {Letter of Intent: The Hyper-Kamiokande Experiment --- Detector
  Design and Physics Potential ---} (2011),  {{arXiv:1109.3262}}.

\bibitem{Aalseth:2017fik}
C.E. Aalseth et~al., Eur. Phys. J. Plus, {\bf 133}, 131 (2018),
  {{arXiv:1707.08145}}.

\bibitem{Aprile:2020vtw}
E.~Aprile et~al., JCAP, {\bf 11}, 031 (2020),  {{arXiv:2007.08796}}.

\bibitem{Agostini:2020adk}
F.~Agostini et~al., Eur. Phys. J. C, {\bf 80}, 808 (2020),
  {{arXiv:2003.13407}}.

\bibitem{Simgen:2003et}
H.~Simgen, C.~Buck, G.~Heusser, M.~Laubenstein, and W.~Rau, Nucl. Instrum.
  Meth. A, {\bf 497}, 407--413 (2003).

\bibitem{Yu:2020vbz}
X.~Yu et~al., JINST, {\bf 15}, P09001 (2020).

\bibitem{hoppe2008}
E.~Hoppe, C.~Aalseth, R.~Brodzinski, A.~Day, O.~Farmer, T.~Hossbach,
  J.~McIntyre, H.~Miley, E.~Mintzer, A.~Seifert, J.~Smart, and G.~Warren, J.
  Radioanal. Nucl. Chem., {\bf 277}, 103--110 (2008).

\bibitem{Alduino:2016vjd}
C.~Alduino et~al., JINST, {\bf 11}, P07009 (2016),  {{arXiv:1604.05465}}.

\bibitem{Abe:2017jzw}
K.~Abe et~al., Nucl. Instrum. Meth. A, {\bf 884}, 157--161 (2018),
  {{arXiv:1707.06413}}.

\bibitem{Bunker:2020sxw}
R.~Bunker, T.~Aramaki, I.J. Arnquist, R.~Calkins, J.~Cooley, E.W. Hoppe, J.L.
  Orrell, and K.S. Thommasson, Nucl. Instrum. Meth. A, {\bf 967}, 163870
  (2020),  {{arXiv:2003.06357}}.

\bibitem{Hoppe:2014nva}
E.W. Hoppe, C.E. Aalseth, O.T. Farmer, T.W. Hossbach, M.~Liezers, H.S. Miley,
  N.R. Overman, and J.H. Reeves, Nucl. Instrum. Meth. A, {\bf 764}, 116--121
  (2014).

\bibitem{Balogh:2020nmo}
L.~Balogh et~al., Nucl. Instrum. Meth. A, {\bf 988}, 164844 (2021),
  {{arXiv:2008.03153}}.

\bibitem{2011JInst...6C2018F}
A.~{Fr{\"o}jdh}, G.~{Thungstr{\"o}m}, C.~{Fr{\"o}jdh}, and S.~{Petersson},
  JINST, {\bf 6}, C12018 (2011).

\bibitem{osti_5305900}
T.~Iida, Y.~Ikebe, T.~Hattori, H.~Yamanishi, S.~Abe, K.~Ochifuji, and
  S.~Yokoyama, Health Phys., {\bf 54} (1988).

\bibitem{howard1991}
A.J. Howard, S.E. Carroll, and W.P. Strange, Am. J. Phys., {\bf 59}, 544--550
  (1991).

\bibitem{osti_6605848}
K.D. Chu and P.K. Hopke, Environ. Sci. Technol., {\bf 22} (1988).

\bibitem{takeda:nu2020}
A.~Takeda, K.~Okamoto, S.~Yamamoto, Y.~Takeuchi, and Y.~Nakano,
\newblock Development of high-sensitivity radon detector in water with
  continuous measurement (2020),
\newblock Poster presentation at the XXIX International Conference on Neutrino
  Physics and Astrophysics.

\bibitem{Leonard:2007uv}
D.S. Leonard et~al., Nucl. Instrum. Meth. A, {\bf 591}, 490--509 (2008),
  {{arXiv:0709.4524}}.

\bibitem{Aprile:2011ru}
E.~Aprile et~al., Astropart. Phys., {\bf 35}, 43--49 (2011),
  {{arXiv:1103.5831}}.

\bibitem{Wang:2016eud}
X.~Wang et~al., JINST, {\bf 11}, T12002 (2016),  {{arXiv:1608.08345}}.

\bibitem{Aprile:2017ilq}
E.~Aprile et~al., Eur. Phys. J. C, {\bf 77}, 890 (2017),  {{arXiv:1705.01828}}.

\bibitem{Cebrian:2017jzb}
S.~Cebri\'an et~al., JINST, {\bf 12}, T08003 (2017),  {{arXiv:1706.06012}}.

\bibitem{Sekiya:2015rla}
H.~Sekiya, AIP Conf. Proc., {\bf 1672}, 080001 (2015).

\bibitem{Takeuchi:2017pij}
Y.~Takeuchi, S.~Tasaka, and Y.~Nakano, J. Phys. Conf. Ser., {\bf 888}, 012211
  (2017).

\end{thebibliography}
%

\appendix

\section{Fitting parameters of empirical function} \label{sec_app1}
As briefly explained in Sect.~\ref{sec_ah}, the behavior of the absolute humidity dependence of the calibration factor is well described by an empirical function, $C_{F}=A-B\sqrt{A_{H}}$. Table~\ref{table_fit} summarizes the fitting parameters obtained in this study as well as the past studies.

\begin{table*}
\centering
{\scriptsize
\caption{Summary of the fitting parameters of the empirical function. Those parameters are obtained by filling purified air.}
\label{table_fit}
\begin{tabular}{cccccc}
\hline
Detector & Size of  & High voltage & $A$ & $B$ & Comment \\ 
Volume & PIN-photodiode & [kV] & & & \\
\hline
$1$~L & $10\times10~\mathrm{mm}$ & $-0.12$ & $(12.86\pm0.40)\times10^{-4}$ & $(1.66\pm0.19)\times10^{-4}$ & Ref.~\cite{Pronost:2018ghn} \\
$70$~L & $16\times16~\mathrm{mm}$ & $-1.5$ & $2.03$ & $1.08$ & Ref.~\cite{TAKEUCHI1999334} \\
 & & & & & Constant~(0.73)  \\
  &  & & & & above $1.6~\mathrm{g/m^{3}}$ \\
$80$~L & $18\times18~\mathrm{mm}$ & $-2.0$ & $2.25\pm0.05$ & $0.29\pm0.05$ & Ref.~\cite{Nakano:2017rsy} \\
$80$~L & $18\times18~\mathrm{mm}$ & $-2.0$ & $2.39\pm0.04$ & $0.42\pm0.05$ & This study \\
$80$~L & $28\times28~\mathrm{mm}$ & $-2.0$ & $2.48\pm0.04$ & $0.37\pm0.05$ & This study \\
\hline
\end{tabular}
}
\end{table*}

\section{Evaluation of the background rate including the calibration setup} \label{sec_bg_setup}

\begin{figure*}
\begin{minipage}{0.5\hsize}
\centering\includegraphics[width=1.0\linewidth]{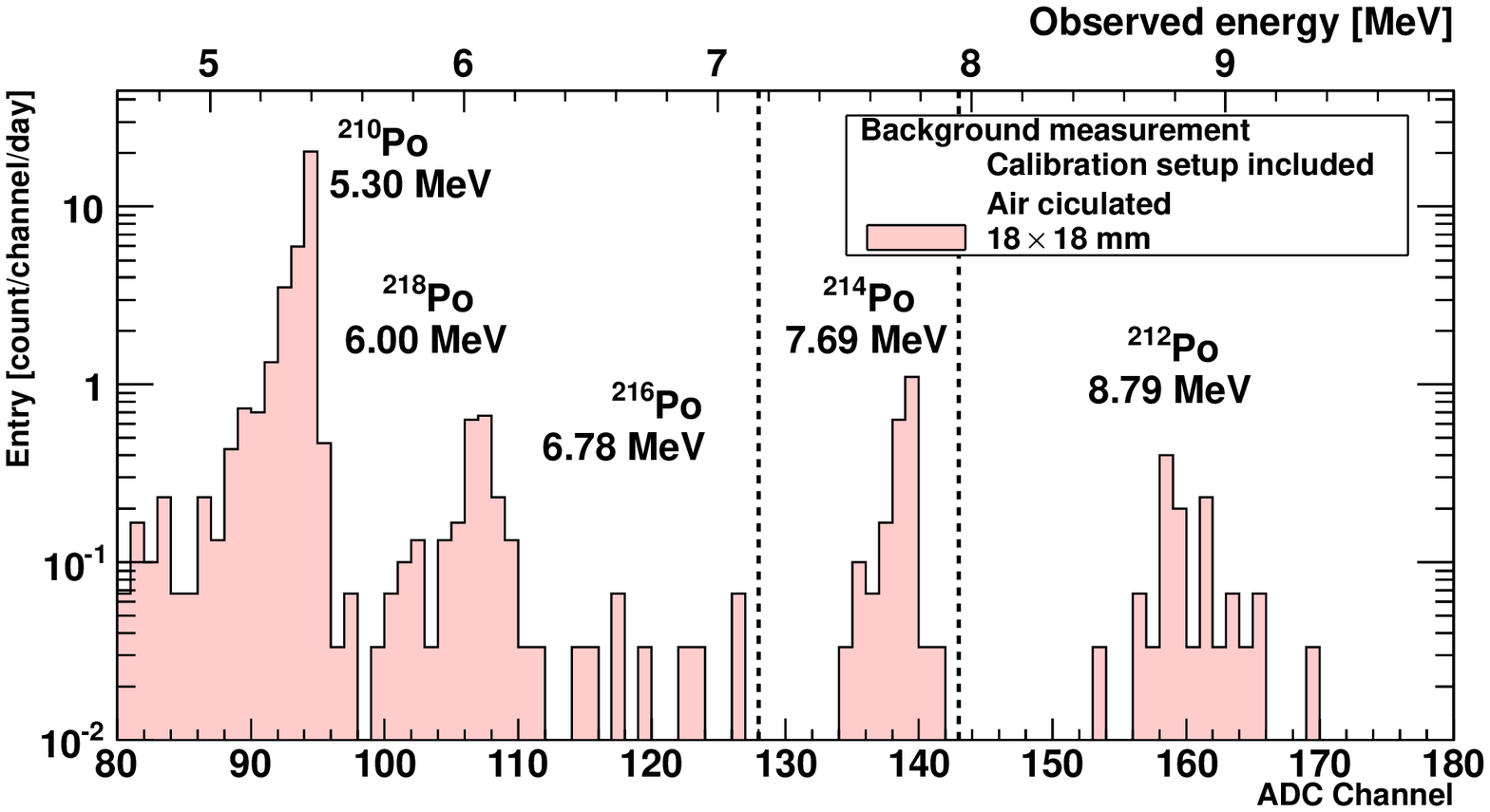}
\end{minipage}
\begin{minipage}{0.5\hsize}
\centering\includegraphics[width=1.0\linewidth]{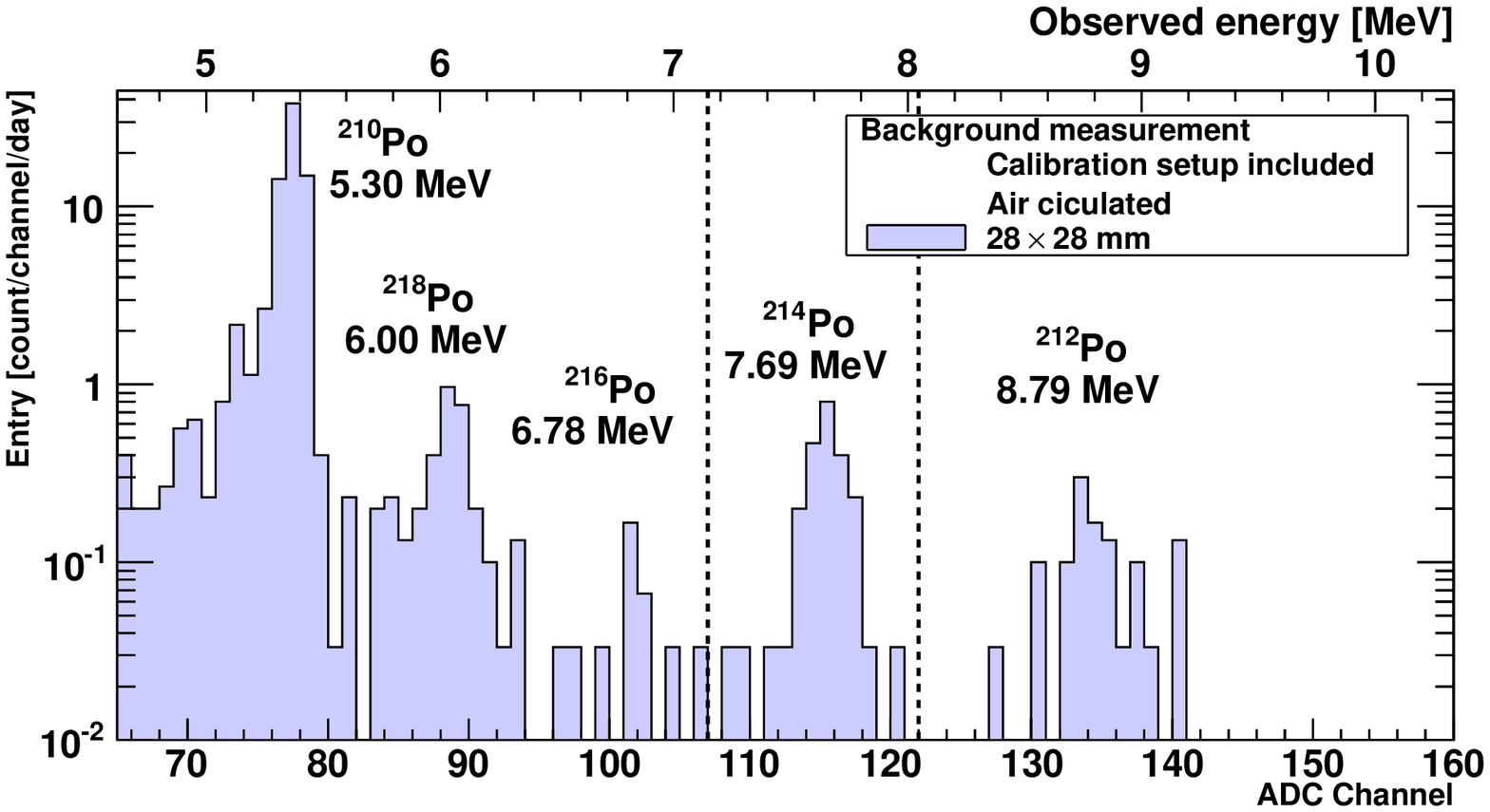}
\end{minipage}
\caption{The daily background spectrum measured by two Rn detectors by circulating the purified air. The left~(right) panel shows the spectrum measured by the Rn detector equipped with $18\times18~\mathrm{mm}$ ~($28\times28~\mathrm{mm}$) PIN-photodiode.}
\label{fig_bg_spect2}
\end{figure*}

We measured the background rate by circulating the purified air in the calibration setup in order to evaluate some contributions of Rn emanation from the other components of the calibration setup.  Figure~\ref{fig_bg_spect2} shows the measured background spectrum when the filled purified air is circulated. The air flow rate was controlled at $0.6$~L/min by the mass flow controller.

\begin{figure*}
\begin{minipage}{0.5\hsize}
\centering\includegraphics[width=1.0\linewidth]{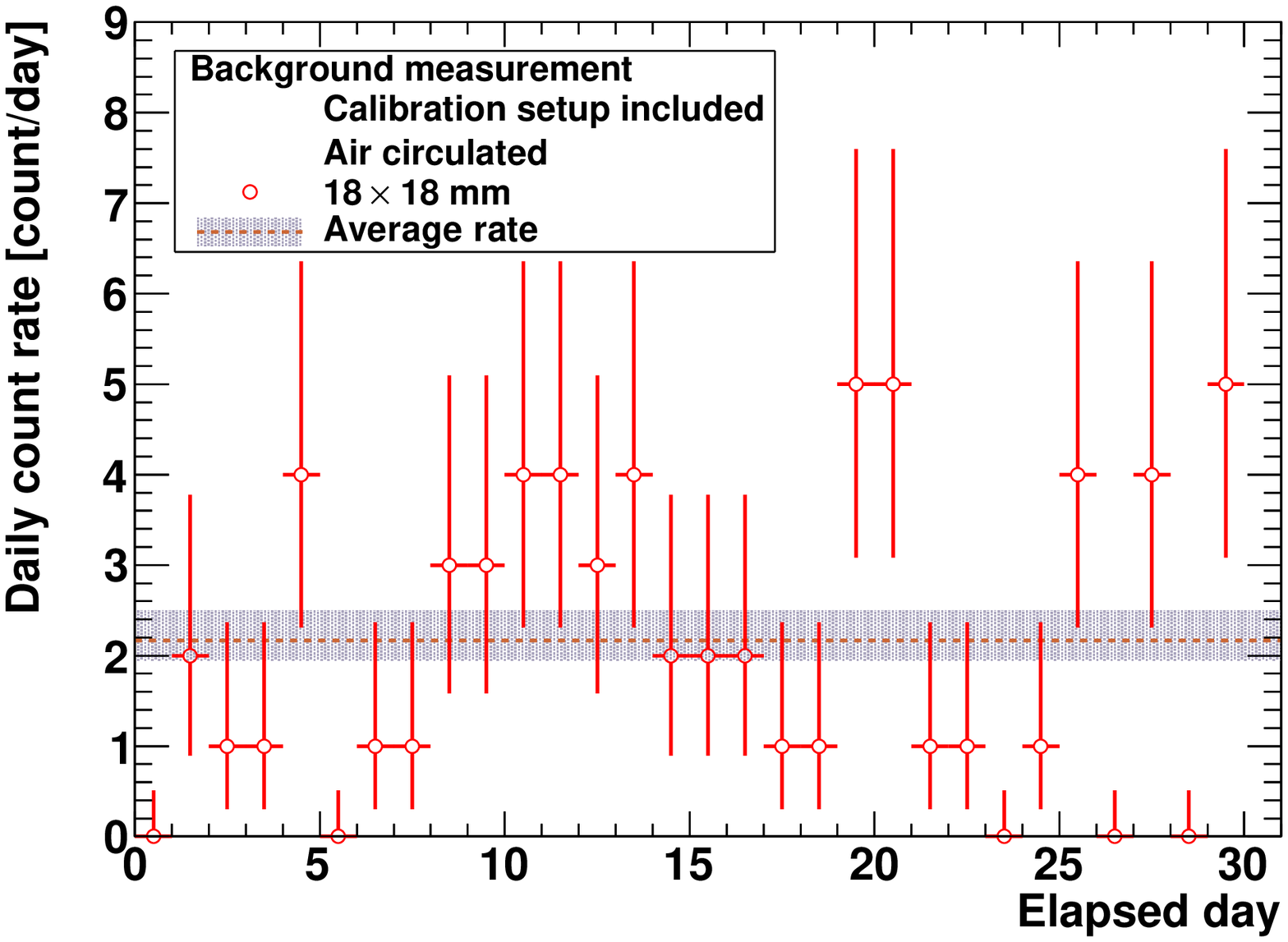}
\end{minipage}
\begin{minipage}{0.5\hsize}
\centering\includegraphics[width=1.0\linewidth]{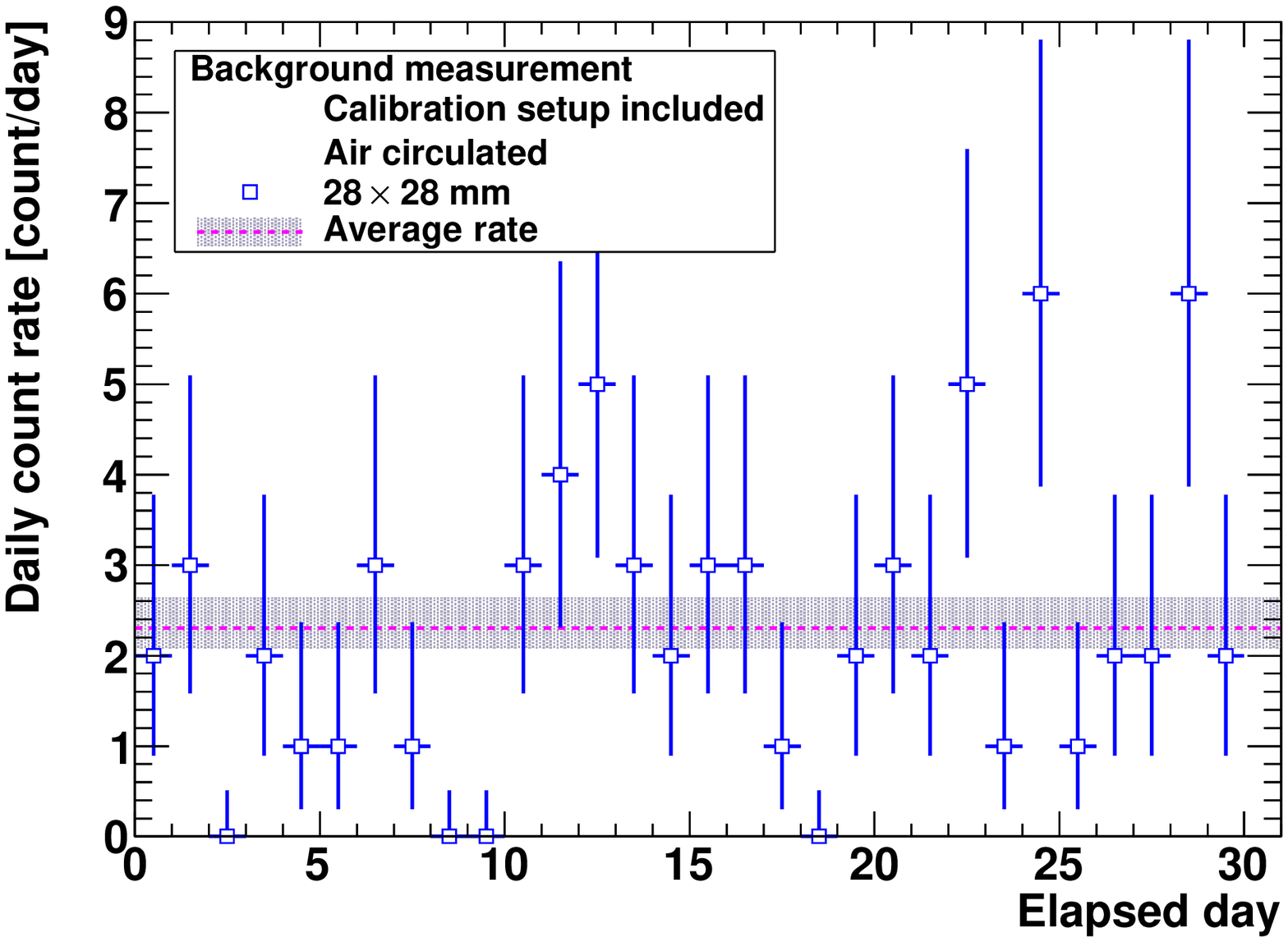}
\end{minipage}
\caption{The measured background rate of the Rn detectors as a function of time. The right (left) panel shows the result from the Rn detector with $18\times18~\mathrm{mm}$ ~($28\times28~\mathrm{mm}$) PIN-photodiode. The horizontal dashed line~(gray band) shows the average~(uncertainty) of background rate for each detector.}
\label{fig_bg_rate2}
\end{figure*}

Figure~\ref{fig_bg_rate2} shows the daily count rate in this measurement. The measured background rate of Rn detector with the large PIN-photodiode is $2.32^{+0.34}_{-0.23}$~count/day while that of Rn detector with the small PIN-photodiode is $2.17^{+0.33}_{-0.22}$~count/day as summarized in Table~\ref{table_bg_setup}.

\begin{table*}
\centering
\caption{Summary of the background rates of the calibration setup. We took $30$~days data for this measurement. The count rate shows the actual count rate including the intrinsic background from the two Rn detectors while the Rn concentration is obtained after subtracting the intrinsic backgrounds described in Sect.~\ref{sec_bg}.}
\label{table_bg_setup}
\begin{tabular}{ccc}
\hline
 &  Rn detector  & Rn detector  \\ 
 & with $28\times28~\mathrm{mm}$ & with $18\times18~\mathrm{mm}$ \\
\hline
Count rate~[count/day]& $2.32^{+0.34}_{-0.23}$ & $2.17^{+0.33}_{-0.22}$ \\ 
(Including intrinsic background & & \\
from two Rn detectors) & & \\ \hline
Rn concentration~[$\mathrm{mBq/m^{3}}$] & $0.40^{+0.19}_{-0.13}$ & $0.38^{+0.17}_{-0.13}$ \\ 
(Intrinsic backgrounds subtracted) & & \\
\hline
\end{tabular}
\end{table*}

The background rates measured by two Rn detectors include their intrinsic backgrounds. Thus, the background from the calibration setup should be obtained by subtracted their intrinsic backgrounds from the measured background rates. For the Rn detector with $28\times28~\mathrm{mm}$ PIN-photodiode, the intrinsic background is the measurement result described in Sect.~\ref{sec_bg}. On the other hand, we have not performed the background measurement for the Rn detector with $18\times18~\mathrm{mm}$ PIN-photodiode in this study. To compensate for this deficiency of background data, we used the past result from Ref.~\cite{Nakano:2017rsy} to subtract the background rate. Finally, the measured Rn concentrations are $0.40^{+0.19}_{-0.13}~\mathrm{mBq/m^{3}}$ by $28\times28~\mathrm{mm}$ and $0.38^{+0.17}_{-0.13}~\mathrm{mBq/m^{3}}$ by $18\times18~\mathrm{mm}$ after subtracting the intrinsic backgrounds of two Rn detectors.

Using the correction factor~$\varepsilon$ defined in Eq.~(\ref{eq_3}), the daily Rn emanation rate is also estimated to be $0.04^{+0.03}_{-0.02}~\mathrm{\mu Bq/setup/day}$. Although some excesses over the intrinsic background are observed, the Rn emanation form the calibration setup is quit low. Moreover, there are no air leaks in the calibration setup since the environmental Rn concentration in the experimental site is about $40$--$80~\mathrm{Bq/m^{3}}$, which has been continuously monitored by the $1$~L Rn detector~\cite{Pronost:2018ghn}. Therefore, the background from the calibration setup does not affect the calibration results described in Sect.~\ref{sec_setup} because of its small contamination.

\end{document}